\documentclass{ws-ijmpa}
\usepackage[super,compress]{cite}
\usepackage{graphicx}

\usepackage{bm}% bold math

\newcommand{\surf}{\parallel}
\newcommand{\dlim}{\displaystyle\lim}
\newcommand{\dint}{\displaystyle\int}
\newcommand{\dsum}{\displaystyle\sum}

\newcommand{\abs}[1]{\vert #1\vert}
\newcommand{\Det}[1]{\left\vert #1\right\vert}

\newcommand{\dd}{\mathrm{d}}
\newcommand{\ii}{\mathrm{i}}
\newcommand{\Eq}[1]{Eq.~\eqref{#1}}

\begin{document}
\markboth{Authors' Names}{Instructions for typing manuscripts (paper's title)}

\catchline{}{}{}{}{}
%
%%%%%%%%%%%%%%%%%%%%%%%%%%%%%%%%%%%%%%%%%%%%%%%%%%%%%%%%%%%%%%%%%%%%

\title{Casimir energy between carbon nanotubes}

\author{Pablo Rodriguez-Lopez}

\address{{\'A}rea de Electromagnetismo and Grupo Interdisciplinar de Sistemas Complejos (GISC), Universidad Rey Juan Carlos\\
M{\'o}stoles, Madrid, 28933, Spain\\
pablo.ropez@urjc.es}

\maketitle

\begin{history}
\received{Day Month Year}
\revised{Day Month Year}
\end{history}

\begin{abstract}
We obtain the large distance limit of the Casimir energy between two equal parallel straight single wall carbon nanotubes by the use of the Multiscattering formalism, for low and high temperatures. 
\keywords{Casimir effect; Multiscattering formalism; carbon nanotube.}
\end{abstract}

\ccode{PACS numbers:}

%\tableofcontents

\section{Introduction}
The quantum-thermal fluctuations of the electromagnetic field cause the appeareance of forces between neutral objects, as was predicted by Casimir \cite{Casimir}. Later, Lifshitz generalized the study for finite temperature plates\cite{Lifshitz}. Today, the Lifshitz formula has been generalized to objects with arbitrary geometry by the Multiscattering formalism \cite{MS01}\cite{MS02}\cite{MS03}\cite{MS04}\cite{MS05}\cite{MS06}.

The Casimir force is relevant in the micromechanical range of scales, therefore, it is expected that it can have an important effect in the interaction between between single wall carbon nanotubes (SWCNs). In the large distance distance limit, the Casimir interaction has been computed \cite{Casimir_SWCN01}, also, the Casimir force of one SWCN inside another was obtained in \cite{Casimir_SWCN02}. By using the modern Multiscattering formalism\cite{MS03}, the Casimir energy between parallel cylinders can be easily and sistematically obtained \cite{MS03}, even the case of inclined cylinders is obtained with this formalism \cite{Casimir_inclined_Cylinders}.

Here we are going to apply the Multiscattering formalism to compute the large distance limit between parallel SWCNs, and we are going to obtain analytical results for asymptotic Casimir energies in the zero and high temperature limits.

This article is organized as follows. In Sect. \ref{Sect2}, we introduce the Multiscattering formalism that we are going to use for objects with cylindrical symmetry, then we define the cylindrical vector multipoles and the translation and T scattering matrices we are going to use. In Sect. \ref{Sect3} we review the axial electric conductivity of SWCNs, needed to obtain the T-matrix. We show the main results of this article in Sect. \ref{Sect4} and we summarize our results in Sect. \ref{Sect5}.

\section{Formalism for Casimir effect}\label{Sect2}
Here we are going to use the multiscattering formalism for the Casimir effect \cite{MS01}\cite{MS02}\cite{MS03}\cite{MS04}\cite{MS05}\cite{MS06}. The Casimir energy $E_{T}$ between 2 objects at any finite temperature $T$ is calculated by
\begin{eqnarray}
E_{T} = k_{B}T{\dsum_{n=0}^{\infty}}'\log\Det{ \mathbb{I} - \mathbb{T}_{1}(\kappa_{n})\mathbb{U}_{12}(\kappa_{n})\mathbb{T}_{2}(\kappa_{n})\mathbb{U}_{21}(\kappa_{n}) },
\end{eqnarray}
where $\omega = \ii\kappa_{n}$, $\kappa_{n} = 2\pi\tfrac{k_{B}T}{\hbar c}n$ are the Matsubara frequencies, $k_{B}$ is the Boltzmann constant, $\hbar$ the Planck constant, $c$ the light velocity, $\mathbb{I}$ is the identity matrix, $\mathbb{T}_{a}$ is the T-scattering matrix of the object $a$ placed at $\bm{x}_{a}$, and $\mathbb{U}_{ab}$ is the translation matrix in open space of multipole waves from $\bm{x}_{a}$ to $\bm{x}_{b}$ and the tilde in the sumatory means that the $n=0$ case has a $1/2$ weight.

In the case of parallel infinite cylinders, all matrices are diagonal in the cylindrical axes (labelled by $z$ here) and the formula is simplified into
\begin{eqnarray}\label{CasimirEnergyT}
E_{T} = k_{B}T{\dsum_{n=0}^{\infty}}'\dint_{-\infty}^{\infty}\dfrac{\dd k_{z}}{\left(\tfrac{2\pi}{L}\right)}\log\Det{ \mathbb{I} - \mathbb{T}_{1}(\kappa_{n})\mathbb{U}_{12}(\kappa_{n})\mathbb{T}_{2}(\kappa_{n})\mathbb{U}_{21}(\kappa_{n}) }
\end{eqnarray}
where $L$ is the lenght of the cylinders. This formula can be simplified in the zero and high temperature (or classical) limit as
\begin{eqnarray}\label{CasimirEnergy0}
E_{0} = \dfrac{\hbar c}{2\pi}\dint_{0}^{\infty}\dd\kappa\dint_{-\infty}^{\infty}\dfrac{\dd k_{z}}{\left(\tfrac{2\pi}{L}\right)}\log\Det{ \mathbb{I} - \mathbb{T}_{1}(\kappa)\mathbb{U}_{12}(\kappa)\mathbb{T}_{2}(\kappa)\mathbb{U}_{21}(\kappa) }
\end{eqnarray}
and
\begin{eqnarray}\label{CasimirEnergycl}
E_{cl} = \dfrac{k_{B}T}{2}\dint_{-\infty}^{\infty}\dfrac{\dd k_{z}}{\left(\tfrac{2\pi}{L}\right)}\log\Det{ \mathbb{I} - \mathbb{T}_{1}(0)\mathbb{U}_{12}(0)\mathbb{T}_{2}(0)\mathbb{U}_{21}(0) }
\end{eqnarray}
respectively.

\subsection{Vector multipoles}
We need a basis to obtain the $\mathbb{T}$ scattering matrices for SWCN and to define the translations matrices $\mathbb{U}$ between the objects. We use the following basis for vector multipoles\cite{MS03}
\begin{eqnarray}
\bm{M}_{n k_{z}}(\bm{x},k) = \dfrac{1}{\abs{\kappa_{\rho}}}\bm{\nabla}\times\big[\phi_{n k_{z}}(\bm{x},k)\hat{\bm{z}} \big]
\end{eqnarray}
\begin{eqnarray}
\bm{N}_{n k_{z}}(\bm{x},k) = \dfrac{1}{\kappa}\bm{\nabla}\times\bm{M}_{n k_{z}}(\bm{x},k)
\end{eqnarray}
for imaginary frequencies $\omega = \ii\xi = \ii c\kappa$ and $\kappa_{\rho} = \sqrt{ \xi^{2} + k_{z}^{2} } \ge 0$, $\phi_{n k_{z}}^{\text{reg}}(\bm{x},\kappa) = I_{n}(\kappa_{\rho}\rho)e^{\ii n\theta}e^{\ii k_{z}z}$ and $\phi_{n k_{z}}^{\text{out}}(\bm{x},\kappa) = K_{n}(\kappa_{\rho}\rho)e^{\ii n\theta}e^{\ii k_{z}z}$.
\subsection{Translation U-matrices}
Using the results of \cite{MS03}, the translation matrix $\mathbb{U}_{ab}$ in open space of multipole waves from $\bm{x}_{a}$ to $\bm{x}_{b}$ is a matrix with the following entries
\begin{eqnarray}
P^{\text{out}}_{nk_{z}}(\bm{x}_{i}) = \int_{-\infty}^{\infty}\dfrac{\dd q_{z}}{\left(\tfrac{2\pi}{L}\right)}\sum_{m\in\mathbb{Z}}\sum_{Q}^{\{\bm{M},\bm{N}\}}\mathcal{U}_{nk_{z}P,mq_{z} Q}^{ij}Q^{\text{reg}}_{mq_{z}}(\bm{x}_{j})
\end{eqnarray}
where $\bm{X}_{ij} = \bm{x}_{i} - \bm{x}_{j} = \big(X^{ij}_{\perp}\cos(\theta^{ij}), X^{ij}_{\perp}\sin(\theta^{ij}),X^{ij}_{z}\big)$, $P$ and $Q$ represent the vectorial multipoles $\bm{M}(\kappa)$ and $\bm{N}(\kappa)$ and
\begin{eqnarray}\label{CylTranslation_matrix}
\mathcal{U}_{nk_{z}P;mq_{z}Q}^{ij} = \ii^{m}K_{n-m}\left(X^{ij}_{\perp}\sqrt{ \kappa^{2} + k_{z}^{2} }\right)e^{-\ii k_{z}X^{ij}_{z}}e^{-\ii(n-m)\theta^{ij}}\delta_{PQ}\dfrac{\delta(k_{z} - q_{z})}{\left(\tfrac{L}{2\pi}\right)}
\end{eqnarray}
for vectorial multipoles defined in systems of coordinates with parallel axes but displaced by $\bm{X}_{ij}$.

\subsection{T-matrix for SWCN}
By using the Waterman formalism \cite{Waterman}, we obtain the T matrix for a cylindrical surface of radius $R$ with only axial conductivity $\sigma_{zz}(\xi)\neq 0$ (for compact dielectric cylinders see \cite{TMatrix_Cylinder}) for imaginary frequencies $\omega = \ii\xi$ is
\begin{eqnarray}\label{T_matrix_SWCN}
\mathcal{T}_{nk_{z}P,mq_{z}Q}(\xi) = \dfrac{(-1)^{n+1}\dfrac{\pi}{2}R\dfrac{\kappa_{\rho}^{2}}{\xi}\sigma_{zz}(\xi,k_{z})I_{n}^{2}(R\kappa_{\rho})}{ 1 + R\dfrac{\kappa_{\rho}^{2}}{\xi}\sigma_{zz}(\xi,k_{z})I_{n}(R\kappa_{\rho})K_{n}(R\kappa_{\rho}) }\delta_{nm}\dfrac{\delta(k_{z} - q_{z})}{\left(\tfrac{L}{2\pi}\right)}\delta_{P,N}\delta_{Q,N}
\end{eqnarray}
This result coincides with \cite{TMatrix_SWCN_Igor}, but differs from \cite{TMatrix_SWCN_Garcia_Abajo} because in our case we are considering that the azimutal conductivity is zero, while in \cite{TMatrix_SWCN_Garcia_Abajo} it is assumed that $\sigma_{\phi\phi} = \sigma_{zz}$. Note that this T matrix depends on the electric axial conductivity of the SWCN, that we are ging to discuss in the following section.

\section{Conductivity of SWCNs}\label{Sect3}
As we are interested in the large distance limit ($d\gg R$) of the Casimir energy between carbon nanotubes, we only need to know the behavior of the conductivity in a small neighborhood around $\omega = 0$. For those small frequencies, the azimutal conductivity is much smaller than the axial conductivity $\sigma_{\phi\phi} \ll\sigma_{zz}$, then, we can safety approach $\sigma_{\phi\phi} = 0$ in our analysis \cite{Casimir-SWCN01}. In addition to that, for very small frequencies the contribution of exitonic peaks is neglitible \cite{Casimir-SWCN01}\cite{Casimir_SWCN02}. Therefore, we characterize the carbon nanotubes only by their axial conductivity $\sigma_{zz}(\omega)$.

The basic properties of each SWCN are captured by their chirality index $(n, m)$, which also determines the SWCN radius $R = \frac{\sqrt{3}a}{2\pi}\sqrt{ m^{2} + nm + n^{2} }$ ($a = 1.42 \text{ \AA}$ is the $C-C$ interatomic distance) and the electronic modulation index $\nu = \{-1,0,+1\}$. The Hamiltonian for the electronic quasiparticle with spin $s$ of a SWCN is approximated by \cite{sigma-SWCN01}
\begin{eqnarray}
\hat{H}_{s}(\bm{K}_{\eta} + \bm{k}_{\surf}) = \hat{H}_{s}^{\eta}(\bm{k}_{\surf}) = \hbar v_{F}\left( \eta k_{n}^{\nu}\tau_{1} + k_{2}\tau_{2} \right) + \Delta_{s}^{\eta}\tau_{3}.
%- \mu\tau_{0},
\end{eqnarray}
where $\tau_{i}$ is the $i^{\underline{th}}$ Pauli matrix of the sublattice pseudo-spin for the A and B sites, and $\eta=\pm 1$ is the valley index, and $k_{n}^{\nu} = \dfrac{1}{R}\left( n - \dfrac{\nu}{3} \right)$ with $n\in\mathbb{Z}$ and $\nu = \{-1,0,+1\}$ is the quantized azimutal momentum due to the cylindrical symmetry of the SWCN. The effective Dirac mass of the line is given by $\Delta_{n\nu}^{\eta s} = \sqrt{ (\Delta_{s}^{\eta})^{2} + (\hbar v_{F}k_{n}^{\nu})^{2} }$.

The local conductivity $\dlim_{\bm{q}\to\bm{0}}\sigma_{zz}(\xi, \bm{q}) = \sigma_{zz}(\xi)$ of a 1D Dirac hyperbola for imaginary frequencies $\omega = \ii\xi$ at zero temperature and with chemical potential $\mu$ consists on the sum of its intraband and interband contributions
\begin{eqnarray}
\sigma_{zz}(\xi, \Delta_{n\nu}^{\eta s}, \mu) = \sigma_{zz}^{\text{intra}}(\xi, \Delta_{n\nu}^{\eta s}, \mu) + \sigma_{zz}^{\text{inter}}(\xi, \Delta_{n\nu}^{\eta s}, \mu)
\end{eqnarray}
where,  by using the Kubo formula, we obtain
\begin{eqnarray}
\sigma_{zz}^{\text{intra}}(\xi, \Delta, \mu) = \dfrac{\alpha c}{\pi}\dfrac{\hbar v_{F}}{\Xi}\dfrac{\sqrt{ \mu^{2} - \Delta^{2} }}{\abs{\mu}}\Theta\big( \abs{\mu} - \abs{\Delta} \big)
\end{eqnarray}
\begin{eqnarray}
\sigma_{zz}^{\text{inter}}(\xi,\mu,\Delta) = \dfrac{\alpha c}{\pi}\frac{\hbar v_{F}}{\Xi}\left( \dfrac{\text{S}}{\text{M}} - 4\Delta^{2}\dfrac{\tanh^{-1}\left(\frac{ \Xi\text{S}\sqrt{ \Xi^{2} + 4\Delta^{2} } }{ \Xi^{2}\text{S} + 4\text{M}\Delta^{2}}\right)}{\Xi\sqrt{ \Xi^{2} + 4\Delta^{2} }}\right)
\end{eqnarray}
where $\text{S} = \abs{\text{M}} - \sqrt{ \text{M}^{2} - \Delta^{2} }$, $\text{M} = \text{Max}\left[\abs{\mu},\abs{\Delta}\right]$, $\alpha\approx\tfrac{1}{137}$ is the fine structure constant, $c$ the light velocity, $\hbar$ the Planck constant, $v_{F}\approx\tfrac{c}{300}$ the Fermi velocity of electronic quasiparticles in the SWCN, $\Xi = \hbar\xi + \hbar\Gamma$ and $\Gamma = 1/\tau$ are the unavoidable losses of the electronic quasiparticles.

The complete axial conductivity is given by
\begin{eqnarray}
\sigma_{zz}(\xi, \mu) = \sum_{\eta=\pm1}\sum_{s=\pm1}\sum_{n=-N}^{N}\sigma_{zz}(\xi, \Delta_{n\nu}^{\eta s}, \mu),
\end{eqnarray}
where $N$ ($-N$) is the maximum (minimum) allowed index for $k_{n}^{\nu}$.

Note that, when $\Delta=0$, the SWCN is always metallic, because there is at least one band crossed by the chemical potential $\mu$, and that, for very small frequencies, we have
\begin{eqnarray}
\sigma_{zz}^{\text{inter}}(\xi,\Delta) = \dfrac{\alpha c \hbar v_{F}}{\pi}
\frac{\text{S} \left(\text{M}\text{S}+\Delta^{2}\right)}{12\Delta^{2}\text{M}^{3}}
\Xi + \mathcal{O}\left[\Xi^{2}\right]
\end{eqnarray}

For finite temperatures we can apply the Maldague formula \cite{Maldague} to obtain the conductivity, it is easy to see that, for any temperature $T>0$, the intraband terms make $\sigma_{zz}(0,\Delta,\mu) > 0$.

\section{Results}\label{Sect4}
Using the multiscattering formula for the Casimir energy for parallel cylinders given in \Eq{CasimirEnergyT}, the translation matrix given in \Eq{CylTranslation_matrix} and the T matrix given in \Eq{T_matrix_SWCN}, we are going to obtain the large distance behavior of the Casimir energy between SWCNs. As the $\mathcal{T}$ operator is only different from zero for the $NN$ polarization term, the problem is effectively reduced to the study of a scalar Casimir effect. In the large distance limit, only the conductivity at very small frequencies is relevant, the electric conductivity can be expanded as
\begin{eqnarray}
\sigma_{zz}(\xi) = \dfrac{\sigma_{-1}}{\xi} + \sigma_{0} + \sigma_{1}\xi + \mathcal{O}\left[\xi^{2}\right]
\end{eqnarray}
Only in the dissipation-less limit, when $\Gamma = 0$, we have $\sigma_{-1} > 0$, when $T>0$ we have $\sigma_{0} > 0$ and, in the zero temperature limit, we have $\sigma_{0} > 0$ for metallic SWCN with $\Gamma > 0$ and $\sigma_{0} = 0$ and $\sigma_{1} > 0$ for dielectric SWCN. We are going to study each one of those cases separately. In several experiments it has been observed that the use of the plasma model (the use of the conductivity in the dissipationless limit $\Gamma\to0$) provides a much better fit of the experimental results \cite{Plasma_Exp}, therefore, it is worth to study this limiting case here.

Using $r = \tfrac{R}{d}$, the Casimir energy in the large distance limit at $T=0$ for two equal dissipation-less ($\Gamma=0$) SWCNs of the same radius $R$ is
\begin{eqnarray}
\mathcal{E}_{0} = \dfrac{E}{L} = - \dfrac{\hbar c \pi}{384 d^{2}\log^{3/2}(r^{-1})}\sqrt{\dfrac{r\sigma_{-1}}{d}},
\end{eqnarray}
for two equal metalic SWCNs with dissipation ($\Gamma>0$) of the same radius $R$
\begin{eqnarray}
\mathcal{E}_{0} = - \dfrac{\hbar c\pi^{2}r\sigma_{0}}{256 d^{2}\log(r^{-1})},
\end{eqnarray}
and for two equal dielectric SWCNs with of the same radius $R$
\begin{eqnarray}
\mathcal{E}_{0} = - \dfrac{\hbar c\pi r^{2}\sigma_{1}^{2}}{15 d^{4}}.
\end{eqnarray}
The Casimir energy in the large distance limit for two equal metallic SWCNs (With and without dissipation $\Gamma$) of the same radius $R$ is
\begin{eqnarray}\label{universal_highT_limit}
\mathcal{E}_{cl} = \dfrac{E}{L} = - \dfrac{k_{B}T\pi^{3}}{32d\log(r^{-1})},
\end{eqnarray}
as in the high temperature limit the intraband conductivity of dielectric become to contribute to the electric conductivity (something easy to probe using the Maldague formula), the Casimir energy between dielectrics is also given by \Eq{universal_highT_limit}, being that result an universal high T limit for SWCNs.

\section{Conclusions}\label{Sect5}
We have obtained the far distance limit ($d \gg R$) of the Casimir energy between two equal SWCNs for the quantum ($T=0$) and classical limits $(dk_{B}T\ll \hbar c)$. We obtain that, depending of the electronic properties of the SWCNs (dielectric, or metallic with or without dissipation), the power law decay of the energy is modified. In the high Temperature limit, the Casimir energy for any pair of equal SWCNs is given by the universal result shown in \Eq{universal_highT_limit}.

\section*{Acknowledgments}
P. R.-L. acknowledges support from Ministerio de Ciencia e Innovaci\'on (Spain), Agencia Estatal de Investigaci\'on, under project NAUTILUS (PID2022-139524NB-I00), from AYUDA PUENTE, URJC.

%\begin{thebibliography}{000} %for 3 digits

\end{document}